\documentclass[12pt,a4paper,final]{iopart}
\usepackage{iopams}
\usepackage{graphicx}
\usepackage[breaklinks=true,colorlinks=true,linkcolor=blue,urlcolor=blue,citecolor=blue]{hyperref}

\begin{document}

\title[Factoring RF Fractional QHOs]
{Factorization of the Riesz-Feller Fractional Quantum Harmonic Oscillators}

\author{H C Rosu}
\address{IPICYT, Instituto Potosino de Investigacion Cientifica y Tecnologica,\\
Camino a la presa San Jos\'e 2055, Col. Lomas 4a Secci\'on, 78216 San Luis Potos\'{\i}, S.L.P., Mexico}
%\address{$^2$Address Two, Neverland}
\ead{hcr@ipicyt.edu.mx http://orcid.org/0000-0001-5909-1945}

\author{S C Mancas}
\address{Department of Mathematics, Embry-Riddle Aeronautical University, Daytona Beach, FL 32114-3900, USA}
\ead{mancass@erau.edu http://orcid.org/0000-0003-1175-6869}

%\author[cor1]{Author Three}
%\address{Address Four, Neverland}
%\eads{\mailto{author.three@mail.com}, \mailto{author.three@gmail.com}}

\begin{abstract}
Using the Riesz-Feller fractional derivative, we apply the factorization algorithm to the fractional quantum harmonic oscillator along the lines
previously proposed by Olivar-Romero and Rosas-Ortiz, extending their results. We solve the non-Hermitian fractional eigenvalue problem in the $k$ space by introducing in that space a new class of Hermite `polynomials' that we call Riesz-Feller Hermite `polynomials'. Using the inverse Fourier transform in Mathematica, interesting analytic results for the same eigenvalue problem in the $x$ space are also obtained. Additionally, a more general factorization with two different L\'evy indices is briefly introduced.
%(2014 \textit{J. Phys. A: Math and Theor.} {\bf xx} yyyyyy).

\end{abstract}

\vspace{2pc}
\noindent{\it Keywords}: factorization, fractional, Riesz-Feller, sub-Gaussian
%\maketitle

\section{Introduction}

A type of fractional quantum harmonic oscillator has been first discussed by Laskin in one of his breakthrough papers \cite{ffr0} on fractional quantum mechanics, but he tackled only a semiclassical approximation. Since then, several authors have dealt with the spatial fractional Schr\"odinger equation with different types of fractional derivatives and various potentials presenting contradictory results and arguments \cite{a,b,c,d,e,f}.
%\textcolor{red}{References to be added.}

Some years ago, Olivar-Romero and Rosas-Ortiz \cite{fff1} were first ones to apply the factorization method \cite{mro,m84} to a fractional differential equation choosing precisely the fractional quantum harmonic oscillator as the case study for their considerations. In line with Laskin, they used the Riesz fractional derivative reporting some interesting results and making suggestions for future work.
This motivated us to proceed with a substantial extension of their results, which we present in this paper. In Section 2, we briefly review the factorization method for the standard quantum harmonic oscillator. In Section 3, where the main results of this work can be found, we present the factorization algorithm for the fractional quantum harmonic oscillator with Riesz-Feller derivatives instead of the Riesz ones as employed in~\cite{fff1}. We have been encouraged to work with the non-Hermitian Riesz-Feller kinetic energy in the Hamiltonian for this case by the recent physical results reported by Berman and Moiseyev \cite{bm18} for the same type of Hamiltonian in the case of impenetrable rectangular potential. In Section 4, we briefly address the factorization with different fractional indices, and we end up stating the conclusions of this work.

\section{Factorization of the standard quantum harmonic oscillator revisited}
Setting $\hbar=m=\omega_0=1$, the eigenvalue problem for the standard Hamiltonian operator of the quantum harmonic oscillator is
\begin{equation} \label{e1}
H_{h.o.}\psi_n\equiv\left(-\frac{1}{2}\frac{d^2}{dx^2}+\frac{1}{2}x^2\right)\psi_n=\lambda_n\psi_n~, \qquad n=0,1,2\dots~,
\end{equation}
where $\lambda_n$ is the spectral parameter (dimensionless energy),
is a basic quantum mechanical eigenvalue problems.
Suppose we take $\lambda_0=\epsilon$, where $\epsilon$ is a constant to be specified later. Then, as will be shown next, $\lambda_1=1+\epsilon$, $\lambda_2=2+\epsilon$,..., $\lambda_n=n+\epsilon$, ..., i.e., for a given $\lambda>\lambda_0$, the nearest spectral neighbors from below and from above are $\lambda-1$ and $\lambda+1$, respectively, and one can write (\ref{e1}) in the form
\begin{equation} \label{e1b}
H_{h.o.}\psi=\lambda \psi~.
\end{equation}

By means of the factoring operators
\begin{eqnarray}
& a_1=\frac{1}{\sqrt{2}}\left(-\frac{d}{dx}+x\right)=-\frac{1}{\sqrt{2}}e^{\frac{x^2}{2}}\frac{d}{dx}e^{-\frac{x^2}{2}}  \\%\equiv A^{\dagger}
& a_2=\frac{1}{\sqrt{2}}\left(\frac{d}{dx}+x\right)=\frac{1}{\sqrt{2}}e^{-\frac{x^2}{2}}\frac{d}{dx}e^{\frac{x^2}{2}}~,
\end{eqnarray}
the Hamiltonian $H_{h.o.}$ can be expressed as
\begin{equation} \label{e2}
H_{h.o.}=a_1a_2+\epsilon=a^\dagger a+\epsilon~,
\end{equation}
where $\epsilon$ is a factorization reminder known as the factorization constant, which for the quantum harmonic oscillator is
$\epsilon=1/2$, while the commutator of the factoring operators is $[a_2,a_1]=2\epsilon=1$.
These factoring operators have been introduced in quantum mechanics by Fock and Dirac already in the 1930's,
but as complex conjugated expressions of $a_1$ and $a_2$ called creation and annihilation operators, respectively,
$$
a^\dagger=\frac{1}{\sqrt{2}}\left(x-ip\right)\equiv a_1 \qquad a=\frac{1}{\sqrt{2}}\left(x+ip\right)\equiv a_2~, %\nonumber\\
$$
where $p=-i \frac{d}{dx}$ is the quantum mechanical momentum (recall that $\hbar =1$).
Then, $a_1a_2=H_{h.o.}-\frac{1}{2}$ and $a_2a_1=H_{h.o.}+\frac{1}{2}$.
Hence, the following intertwinning formulas
\begin{equation} \label{e3}
H_{h.o.}a_1=a_1(H_{h.o.}+1)~, \qquad H_{h.o.}a_2=a_2(H_{h.o.}-1)
\end{equation}
allow an algebraic solution method (factorization algorithm) for the eigenvalue problem (\ref{e1b}).
If $\psi$ is an eigenfunction for eigenvalue $\lambda$, then the intertwinning relationships show that $a_1\psi$ and $a_2\psi$ are
the neighbor eigenfunctions at $\lambda+1$ and $\lambda-1$, respectively. The first step of the algorithm is to find the ground state
eigenfunction from the kernel of $a_2$,
$$
a_2\psi_0=0, \quad \frac{d}{dx}e^{\frac{x^2}{2}}\psi_0=0~, \quad \psi_0=N_0e^{-\frac{x^2}{2}}
$$
for which $\lambda_0=\epsilon=\frac{1}{2}$ as can be checked in (\ref{e1b}). The integration constant $N_0$ is fixed
to $N_0=1/\sqrt[4]{\pi}$ through the normalization condition $\int_{-\infty}^{\infty} |\psi_0|^2 dx=1$.

In the second step, one can find each of the excited eigenfunctions $\psi_n$ by applying $n$ times $a_1$ to $\psi_0$,
\begin{equation} \label{e4}
\psi_n=C_n(a_1)^n\psi_0=N_n(-1)^n\left(e^{\frac{x^2}{2}}\frac{d}{dx}e^{-\frac{x^2}{2}}\right)^n e^{-\frac{x^2}{2}}=N_nH_n(x)e^{-\frac{x^2}{2}}~
\end{equation}
at the corresponding eigenvalues $\lambda_n=n+\frac{1}{2}$, where $H_n(x)$ are the Hermite polynomials,
\begin{equation} \label{e5}
H_n=(-1)^n\left(e^{\frac{x^2}{2}}\frac{d}{dx}e^{-\frac{x^2}{2}}\right)^n e^{-\frac{x^2}{2}}~.
\end{equation}
The normalization constants $N_n$ are given by $N_n=N_0/\sqrt{2^n n!}$, obtained by the normalization conditions for $\psi_n$.

\section{The fractional factorization method}

We follow Olivar-Romero and Rosas-Ortiz and consider a pair of operators $A_{\alpha}$ and $B_{\alpha}$ such that
%.........................
\begin{equation}\label{ff1}
H_{\alpha}\equiv\frac{1}{\alpha}\left(-\frac{d^{\alpha}}{dx^{\alpha}}+x^{2}\right)=B_{\alpha}A_{\alpha}+\epsilon_{\alpha},
\end{equation}
where the factorization remainder $\epsilon_{\alpha}$ can be either a number (as in the conventional factorization)
or a fractional-differential operator. The parameter $\alpha$ defines the fractional order of the derivative and is also known as the L\'evy
stability index because for positive values $\alpha \leq 2$ characterizes the L\'evy stable probability distributions, see \cite{ffr0} for more details.
%which are those distributions having the Gaussian property that sums of big numbers of independent and identically random variables tend to the same distribution (up to scalings) as that of the random variables in the sums.
Usually, in fractional quantum mechanics, one works with $\alpha$ in the interval (1,2] because for $\alpha \leq 1$ the
L\'evy distributions have undefined mean.  %and cannot provide numerical values for the results of quantum measurements.
However, in the factorization method, there is no essential change in the formal results for subunit values of $\alpha$.

The simplest expressions for the factoring operators are
\begin{equation}\label{ff2}
A_{\alpha}=\frac{1}{\sqrt{\alpha}}\left(\frac{d^{\alpha/2}}{dx^{\alpha/2}}+x\right)~ \qquad \mbox{and} \qquad  B_{\alpha}=\frac{1}{\sqrt{\alpha}}\left(-\frac{d^{\alpha/2}}{dx^{\alpha/2}}+x\right)~.
\end{equation}
These are the same as proposed in \cite{fff1} up to the scaling $1/\sqrt{\alpha}$ and provide
\begin{equation}\label{ff3}
B_{\alpha}A_{\alpha} =\frac{1}{\alpha}\left(-\frac{d^{\alpha}}{dx^{\alpha}}-\frac{\alpha}{2}\frac{d^{\alpha/2-1}}{dx^{\alpha/2-1}}+x^{2}\right).
\end{equation}
Comparing (\ref{ff3}) with (\ref{ff1}), one obtains
\begin{equation}\label{e-eps}
\epsilon_{\alpha}=\frac{1}{2}\frac{d^{\alpha/2-1}}{dx^{\alpha/2-1}}~,
\end{equation}
which shows that for $\alpha \neq 2$ the factorization remainder
$\epsilon_{\alpha}$ is a fractional differential derivative of order $\frac{\alpha}{2}-1$
while the case $\alpha=2$ leads to the constant $\epsilon_2 =1/2$ and the factoring operators $A_2$ and $B_2$ reduce
to the usual annihilation and creation operators of the standard harmonic oscillator.

\begin{figure}[htb!]%[x!]
 \centering
\resizebox*{1.17\textwidth}{!}{\rotatebox{0}
{\includegraphics{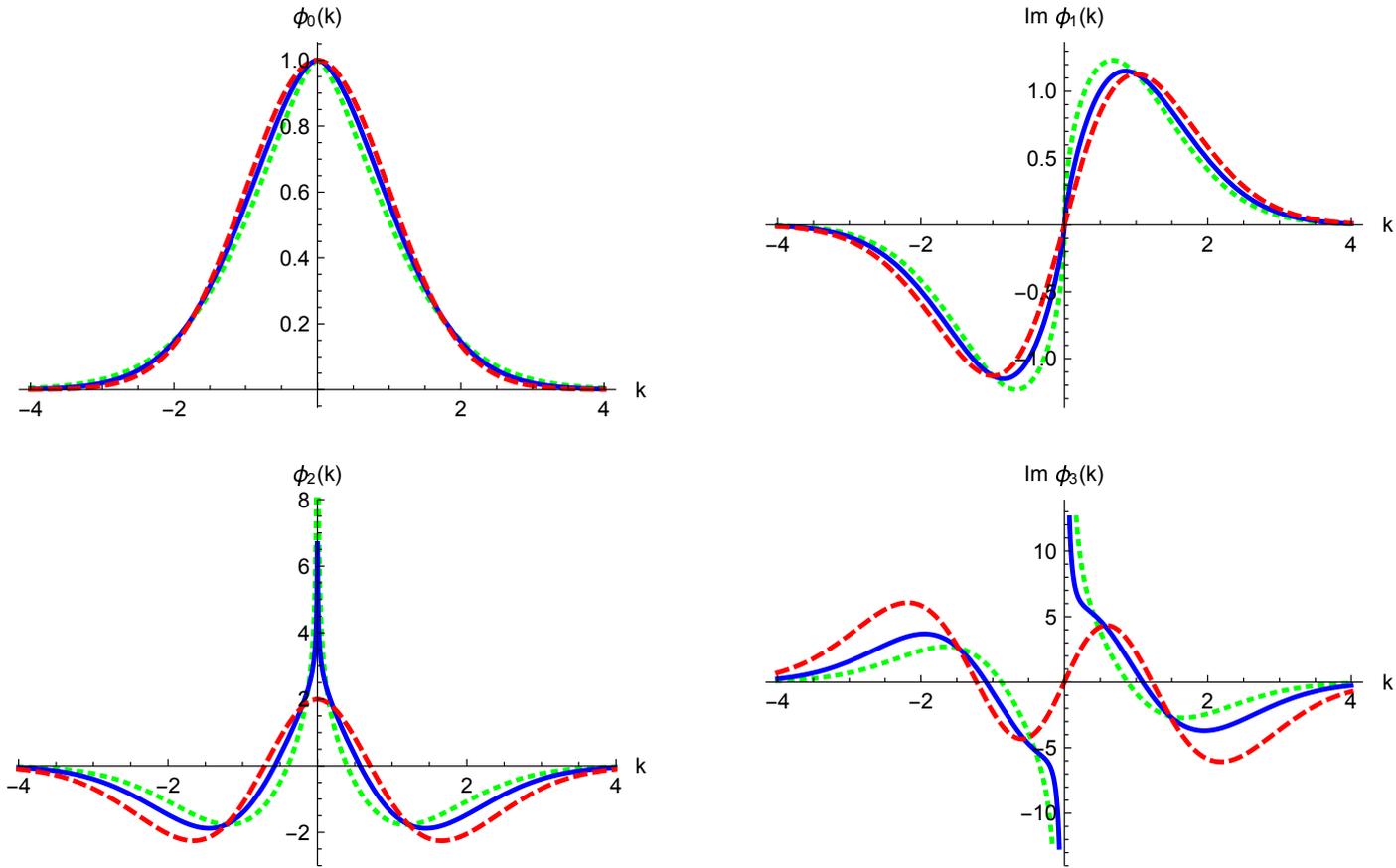}}}
 \caption{\label{fig1} The `ground state' wavefunction and the first three excited wavefunctions in the $k$ space for $\alpha=2$ (red color),
$\alpha=3/2$ (blue color), and $\alpha=1$ (green color). The odd wavefunctions are purely imaginary.}
\end{figure}

\begin{figure}[htb!]%[x!]
 \centering
\resizebox*{1.17\textwidth}{!}{\rotatebox{0}
{\includegraphics{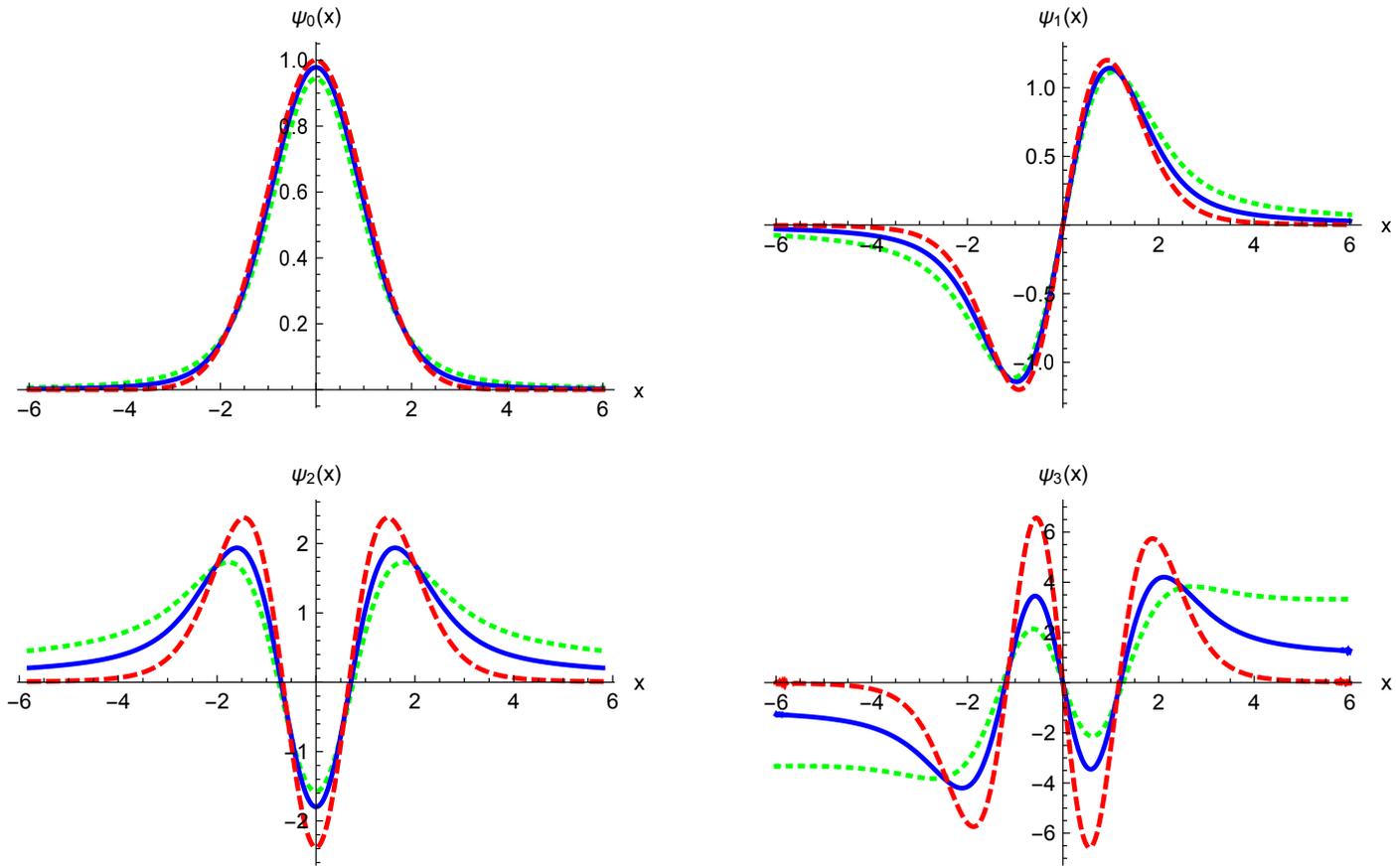}}}
 \caption{\label{fig2} The wavefunctions in the $x$ space obtained by inverse Fourier transforms of the wavefunctions from the previous figure.}
\end{figure}

According to the factorization algorithm, we have to solve the kernel equation of $A_{\alpha}$,
\begin{equation}\label{e-kern}
A_{\alpha}\psi_{0}^{(\alpha)}(x)=0 \quad \longrightarrow \quad \left[\frac{d^{\alpha/2}}{dx^{\alpha/2}}+x\right]
\psi_{0}^{(\alpha)} (x)=0~.
\end{equation}
Since this is a fractional derivative equation, we will solve it in the $k$-space by taking into account that
the Fourier transform $\cal{F}$ of the (quantum) Riesz-Feller derivative $d^\alpha/dx^\alpha$ of a function is characterized by its specific symbol $\Psi_\alpha^\theta$
\begin{equation}\label{e-rf}
{\cal F}\{d^\alpha\psi(x)/dx^\alpha\}:=-
\Psi_\alpha^\theta\,\phi(k)~, \qquad \Psi_\alpha^\theta=|k|^\alpha e^{i\,{\rm sgn}(k)\, \theta \frac{\pi}{2}}~,
\end{equation}
where $\phi(k)={\cal F}\{\psi(x)\}$ and $\theta$ is the skewness (asymmetry) parameter. The latter is usually restricted to numerical values
located at the so-called Takayasu-Feller diamond domain, $|\theta| \leq \, {\rm min}[\alpha,2-\alpha]$ \cite{bm18}.
The factoring operators in the $k$ dual coordinate are
\begin{equation}\label{e-fk}
A_{k,\alpha}=\Psi_{\alpha/2}^{\theta}+i\frac{d}{dk}, \quad \quad \quad \quad B_{k,\alpha}=\Psi_{\alpha/2}^{\theta}-i\frac{d}{dk}
\end{equation}
and the kernel solution of $A_{k,\alpha}$ is the function
\begin{equation}\label{e-phi0}
\phi_{0}^{(\alpha)}(k)=\exp\left(-\frac{|k|^{\alpha/2+1}}{\alpha/2+1}\right)~,
\end{equation}
as shown in Appendix A.
We will call this ground state wavefunction in the dual $k$ coordinate as the fractional sub-Gaussian function for any $\alpha<2$ which turns Gaussian for $\alpha=2$.
All the other excited states are obtained by the repeated usage of the creation operator $B_{k,\alpha}$. For example, the first three excited states
in the $k$ coordinate will be
\begin{eqnarray}\label{e-ex1}
&\phi_{1}^{(\alpha)}(k)=B_{k,\alpha}\phi_0^{(\alpha)}(k)=2i\, {\rm sgn}(k)\
|k|^{\frac{\alpha}{2}}\phi_0^{(\alpha)}~, \\
%\exp\left(-\frac{|k|^{\frac{\alpha}{2}+1}}{\frac{\alpha}{2}+1}\right), \\
&\phi_{2}^{(\alpha)}(k)=B_{k,\alpha}\phi_1^{(\alpha)}(k)=\bigg[\alpha |k|^{\frac{\alpha}{2}-1}-4|k|^\alpha\bigg]\phi_0^{(\alpha)}~,\\
&\phi_{3}^{(\alpha)} (k)=B_{k,\alpha}\phi_2^{(\alpha)}(k)\nonumber \\
&\qquad \quad =i\, {\rm sgn}(k)
\bigg[-8|k|^{\frac{3\alpha}{2}}+6\alpha |k|^{\alpha-1}-
\alpha\left(\frac{\alpha}{2}-1\right)|k|^{\frac{\alpha}{2}-2}\bigg]\phi_0^{(\alpha)}~.
\end{eqnarray}
The eigenfunctions in the $x$ coordinate can be obtained by performing the inverse Fourier transforms of the $\phi$ functions.
In Figs.~\ref{fig1} and ~\ref{fig2}, we present the ground state eigenfunctions and the first three excited eigenfunctions in the $k$ and $x$ coordinates, respectively. All even eigenfunctions $\phi_{2n}$ are real and all the odd eigenfunctions $\phi_{2n+1}$ are purely imaginary, but nevertheless their inverse Fourier transforms, $\psi_{2n+1}$, are real.
%%%%%%%%%%%%%%%%%%%%%%%%%%
%%%%%%%%%%%%%%%%%%%%%%%%%%%

\begin{figure}[htb!]%[x!]
\centering
\resizebox*{0.95\textwidth}{!}{\rotatebox{0}
{\includegraphics{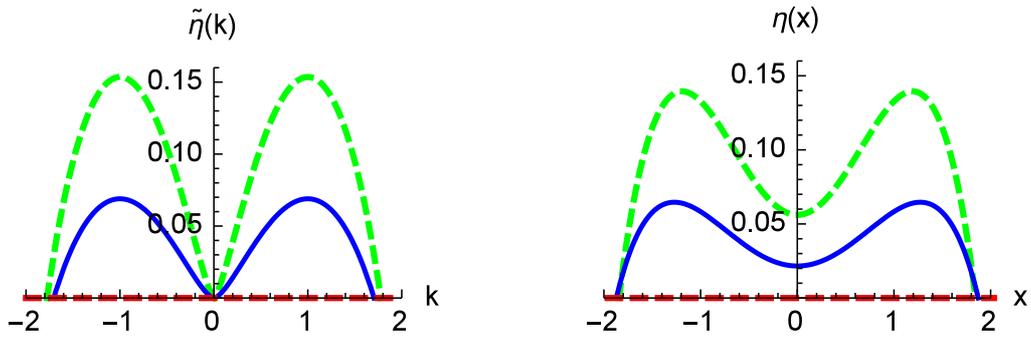}}}
\caption{\label{fig3} Graphs of the non-Gaussian deformations in the $k$ and $x$ spaces for
$\alpha=$1, 3/2 and 2, from top to bottom, respectively.}
\end{figure}

Regarding the bell-shaped $\psi_0^{(\alpha)}$ wavefunctions as obtained from the fractional sub-Gaussians $\phi_0^{(\alpha)}$ for different values of $\alpha$ by the inverse Fourier transform, they can be expressed analytically in terms of a small set of generalized hypergeometric functions according to Mathematica.
The explicit expressions of $\psi_0^{(\alpha)}$ for $\alpha=3/2$ and $\alpha=1$ are provided in Appendix B.
For both $\phi_0^{(\alpha)}$ and $\psi_0^{(\alpha)}$ functions, one can define a degree of non-Gaussianity simply as
\begin{equation}\label{q1}
\tilde{\eta}_\alpha=\frac{\phi_{0}^{(2)}-\phi_{0}^{(\alpha)}}{\phi_{0}^{(2)}}=1-\frac{\phi_{0}^{(\alpha)}}{\phi_{0}^{(2)}}
%\frac{\phi_{\alpha}}{\phi_2}
\qquad {\rm and}\qquad \eta_\alpha=1-\frac{\psi_{0}^{(\alpha)}}{\psi_0^{(2)}},
\end{equation}
respectively. For $\alpha=2$, we have $\tilde{\eta}_2=\eta_2=0$.
It is easy to calculate $\tilde{\eta}_\alpha$ and $\eta_\alpha$ from
\begin{equation}\label{q3}
\tilde{\eta}_\alpha=1-e^{\frac{|k|^{2}}{2}}\phi_{0}^{(\alpha)}
%-\frac{|k|^{\frac{\alpha}{2}+1}}{\frac{\alpha}{2}+1}}
~, \qquad \eta_\alpha=1-e^{\frac{x^2}{2}}\psi_{0}^{(\alpha)}~.
\end{equation}
Both non-Gaussian deformations are displayed in Fig.~\ref{fig3} for the three illustrative values of $\alpha$ used in this paper.
The plots are up to the intersection points with the pure Gaussians, i.e., only for positive $\tilde{\eta}_\alpha$ and $\eta_\alpha$, since for the small negative values in the tail regions there are some numerical problems related to the generalized hypergeometric functions.

\medskip

\begin{figure}[htb!]%[x!]
 \centering
\resizebox*{1.07\textwidth}{!}{\rotatebox{0}
{\includegraphics{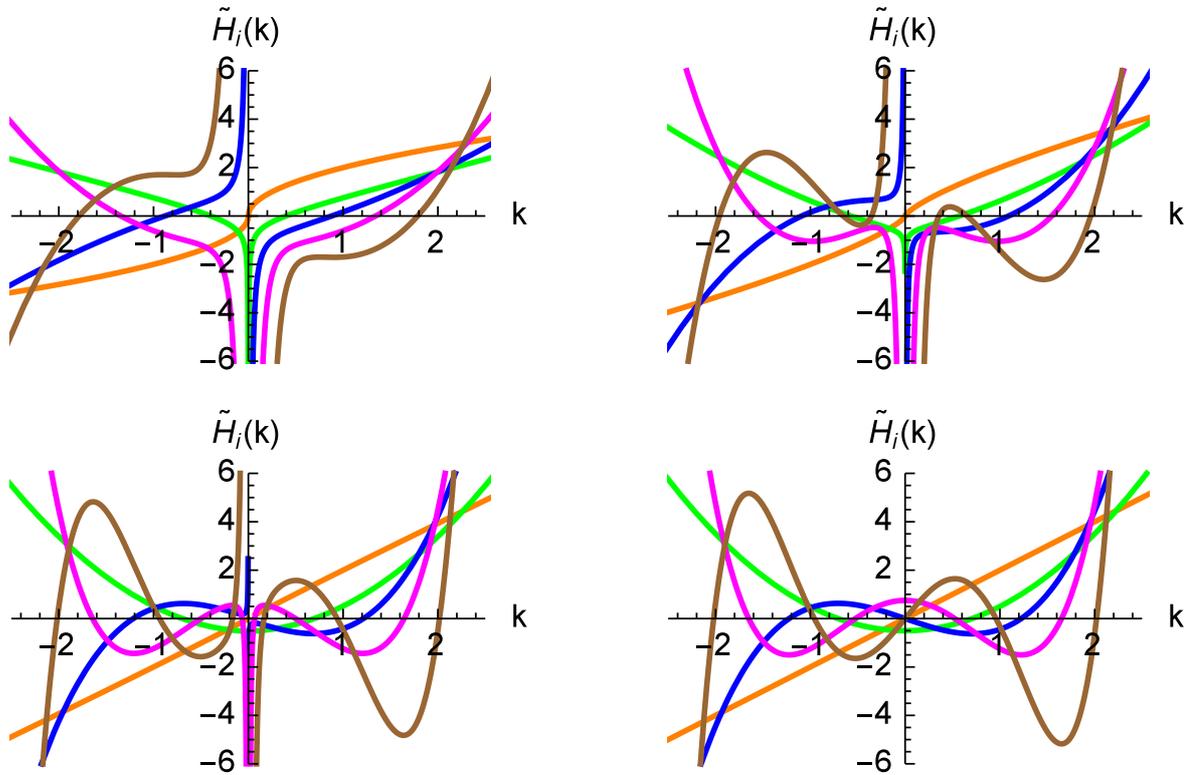}}}
 \caption{\label{fig4} Graphs of the first five Riesz-Feller Hermite `polynomials' scaled by the inverse square of their `degree' in the $k$ space for
$\alpha=$ 1 (top left),1.5 (top right), 1.95 (bottom left), and 2 (bottom right).}
\end{figure}

In general, one can write
\begin{equation}\label{e-phigen}
\phi_n(k)=i^n\widetilde{H}_n\phi_{0}^{(\alpha)}~,
\end{equation}
where $\widetilde{H}_n(k)$ are the fractionally-deformed Hermite `polynomials'
%...............
\begin{eqnarray}
&\widetilde{H}_0=1~ \nonumber\\
&\widetilde{H}_1(k)=2~\mathrm{sgn}(k)|k|^{\frac{\alpha}{2}}\nonumber\\
&\widetilde{H}_2(k)=4|k|^{\frac{2\alpha}{2}}-\alpha|k|^{\frac{\alpha}{2}-1}\nonumber\\
&\widetilde{H}_3(k)=\mathrm{sgn}(k)\bigg[8| k|^{\frac{3\alpha}{2}}-6\alpha|k|^{\frac{2\alpha}{2}-1}+2\frac{\alpha}{2}\left(\frac{\alpha}{2}-1\right)
|k|^{\frac{\alpha}{2}-2}\bigg] \nonumber\\
&\widetilde{H}_4(k)=16|k|^{\frac{4\alpha}{2}}-24\alpha |k|^{\frac{3\alpha}{2}-1}+6\alpha(\alpha-1)|k|^{\frac{2\alpha}{2}-2}\nonumber\\
&+2\frac{\alpha}{2}\left(\frac{\alpha}{2}-1\right)|k|^{\frac{2\alpha}{2}-2}-2\frac{\alpha}{2}\left(\frac{\alpha}{2}-1\right)
\left(\frac{\alpha}{2}-2\right)|k|^{\frac{\alpha}{2}-3} \nonumber\\
&\vdots
\end{eqnarray}
that we also call Riesz-Feller Hermite `polynomials'. For $\alpha=2$, they turn into the standard Hermite polynomials up to a negative sign for the odd ones, though in the $|k|$ variable.
The first five, leaving aside the trivial case of $\widetilde{H}_0$, are plotted in Fig.~\ref{fig4}. Due to the centrifugal type terms (negative powers) present in their expressions for $\alpha<2$, they are singular at the origin unless $\widetilde{H}_1$ which is only discontinuous there.

The general expression for $\widetilde{H}_n(k)$ is
%...........
\begin{eqnarray}\label{genH}
\widetilde{H}_n(k)&=\mathrm{sgn}(k)^n \Big[ 2^n|k|^{\frac{n\alpha}{2}}-p_1(\alpha)|k|^{\frac{(n-1)\alpha}{2}-1}+p_2(\alpha)|k|^{\frac{(n-2)\alpha}{2}-2}\nonumber\\
&-p_3(\alpha)|k|^{\frac{(n-3)\alpha}{2}-3}+...+(-1)^{n-1}p_{n-1}(\alpha)|k|^{\frac{\alpha}{2}-(n-1)}\Big]~,
\end{eqnarray}
where $p_i(\alpha)$ are polynomials of order $i$ in $\alpha$ that can be determined from the following counterpart of the Rodrigues formula
%.................
\begin{equation}\label{rodr}
\widetilde{H}_{n}(k)=
(-1)^n{\rm sgn}(k)^n\,e^{2\frac{|k|^{\frac{\alpha}{2}+1}}{\frac{\alpha}{2}+1}}\frac{d^n}{dk^n}
e^{-2\frac{|k|^{\frac{\alpha}{2}+1}}{\frac{\alpha}{2}+1}}~,
\end{equation}
which for $\alpha=2$, turns into
%.................
\begin{equation}\label{rodr2}
\widetilde{H}_{n}(k)=(-1)^n{\rm sgn}(k)^n\,
e^{|k|^{2}}\frac{d^n}{dk^n}e^{-|k|^{2}}
\end{equation}
to be compared with the standard $x$ space formula
$$
H_{n}(x)=(-1)^n\,
e^{x^{2}}\frac{d^n}{dx^n}e^{-x^{2}}~.
$$
Moving to the calculation of the eigenvalues $\lambda_n(k)$ in the $k$ space, it can be shown that the expressions reported in \cite{fff1} correspond to the asymmetry parameter $\theta=0$, (Riesz derivative). In particular, the first three eigenvalues are
%......eigenvalues
\begin{eqnarray}\label{eig-0}
&\lambda_0(k)=\frac{1}{2}|k|^{\frac{\alpha}{2}-1}~, \\
&\lambda_1(k)=\frac{3}{2}|k|^{\frac{\alpha}{2}-1}-\frac{1}{2}\left(\frac{\alpha}{2}-1\right)|k|^{-2}~, \\
&\lambda_2(k)=\frac{\left(\frac{11\alpha}{2}-6\right)|k|^{\frac{\alpha}{2}-1}-10|k|^\alpha-\left(\frac{\alpha}{2}-1\right)
\left(\frac{\alpha}{2}-2\right)|k|^{-2}}{\alpha-4|k|^{\frac{\alpha}{2}+1}}~.
\end{eqnarray}
However, for $\theta=1$ an additional complex term adds up to each eigenvalue. This term is given by
%.................
\begin{equation}\label{term-even}
\left(\Psi_{\alpha}^{1}-1\right)|k|^{(2n+2)\frac{\alpha}{2}}~, \qquad n=0,1,2,\cdots
\end{equation}
for the even eigenvalues $\lambda_{2n}(k)$ and
%.................
\begin{equation}\label{term-even}
\left(\Psi_{\alpha}^{1}-1\right)|k|^{(2n+1)\frac{\alpha}{2}}~, \qquad n=1,2,\cdots
\end{equation}
for the odd eigenvalues $\lambda_{2n-1}(k)$. This means that all eigenfunctions, including the bell-shaped sub-Gaussians, correspond to
metastable states and we do not expect exceptional points like in the case of impenetrable rectangular wells \cite{bm18}.
%However, we do not discard completely the existence of exceptional points, which remains as an issue for future work.

\section{Factorization with operators of different fractionality}
The use of different L\'evy indices in the factoring operators has been another suggestion in \cite{fff1}.
Although this issue is beyond the scope of this work, we briefly show here how to do it leaving its full consideration for future work.

\medskip

Let us consider the following factoring operators
%..........................
\begin{equation}\label{gd1}
A_\delta=D^{\delta/2}+x~, \qquad B_\gamma=-D^{\gamma/2}+x~,
\end{equation}
where the $D$'s stand for the derivatives of the indicated L\'evy fractional orders.
Then, assuming $\alpha=\frac{\delta+\gamma}{2}$, we obtain
%..........................
\begin{equation}\label{gd4}
{\cal H}_\alpha=B_\gamma A_\delta+\epsilon_{\gamma \delta}~.
\end{equation}
Thus, the remainder operator has the more complicated dissipative form
\begin{equation}\label{gd5}
\epsilon_{\gamma \delta}=\frac{\gamma}{2}D^{\gamma/2-1}+x\left(D^{\gamma/2}-D^{\delta/2}\right)~.
\end{equation}
Notice that $\epsilon_{\gamma\delta}=\epsilon_\alpha$ when $\delta=\gamma\equiv \alpha$.

\medskip

Besides, one can also use the reverted factorization %corresponding to the supersymmetric partner Hamiltonian is
%..........................
\begin{equation}\label{gd2}
{\cal H}_\alpha=-D^\alpha+x^2=A_\delta B_\gamma-\epsilon_{\delta\gamma}~.
\end{equation}
which displays a remainder operator given by
\begin{equation}\label{gd3}
\epsilon_{\delta\gamma}=\frac{\delta}{2}D^{\delta/2-1}+x\left(D^{\delta/2}-D^{\gamma/2}\right)~.
\end{equation}
Notice that the second terms in the two dissipative operators (\ref{gd5}) and (\ref{gd3}) are opposite in sign.

\medskip

Again, for the eigenvalue problems of the factored Hamiltonians, %factored operators $B_\gamma A_\delta$ and $A_\delta B_\gamma$,
one should work in the Fourier $k$-space and come back to the $x$-space by the inverse Fourier transform. To move these operators
%$A_\delta B_\gamma$ operator
in the $k$-space, i.e. to obtain their Fourier counterparts, %$\tilde{A}_\delta \tilde{B}_\gamma$,
the following Fourier transforms are needed
%..................
\begin{eqnarray*}
&{\cal F}\{D^\alpha\psi_0(x)\}=-\Psi_\alpha^\theta\phi_0(k)\nonumber\\
&{\cal F}\{x^2\psi_0(x)\}=-\frac{d^2\phi_0(k)}{dk^2}\nonumber\\
&{\cal F}\{D^{\frac{\delta}{2}-1}\psi_0(x)\}=-\Psi_{\frac{\delta}{2}-1}^{\theta}\phi_0(k) \nonumber\\
&{\cal F}\{xD^{\frac{\delta}{2}}\psi_0(x)\}=i\frac{d}{dk}\left(\Psi_{\frac{\delta}{2}}^{\theta}\phi_0(k)\right)=
i\phi_0(k)\frac{d}{dk}\left(\Psi_{\frac{\delta}{2}}^{\theta}\right)+i\Psi_{\frac{\delta}{2}}^{\theta}\frac{d\phi_0(k)}{dk}~. \nonumber
\end{eqnarray*}

We also need
$$
\frac{d}{dk}\left(\Psi_\alpha^\theta\right)=\frac{d}{dk}\bigg[|k|^\alpha e^{i\,{\rm sgn}(k)\theta\frac{\pi}{2}}\bigg]=\alpha|k|^{\alpha-1}{\rm sgn}(k)
e^{i\,{\rm sgn}(k)\theta\frac{\pi}{2}}=\alpha\,{\rm sgn}(k)\Psi_{\alpha-1}^\theta~.
$$
Using the last two equations, we obtain
$${\cal F}\{xD^{\frac{\delta}{2}}\psi_0(x)\}=
i\frac{\delta}{2}\,{\rm sgn}(k)\Psi_{\frac{\delta}{2}-1}^\theta\phi_0(k)+i\Psi_{\frac{\delta}{2}}^{\theta}\frac{d\phi_0(k)}{dk}~.
$$
Here, we provide the result for the remainder operator $\tilde{\epsilon}_{\gamma\delta}$ in the Fourier space
$$
\tilde{\epsilon}_{\gamma\delta}=-\frac{\gamma}{2}\Psi_{\frac{\gamma}{2}-1}^\theta
+i\bigg[\left(\frac{\gamma\, {\rm sgn}(k)}{2}\Psi_{\frac{\gamma}{2}-1}^\theta+
\Psi_{\frac{\gamma}{2}}^\theta\frac{d}{dk}\right)-\left(\frac{\delta\, {\rm sgn}(k)}{2}\Psi_{\frac{\delta}{2}-1}^\theta+
\Psi_{\frac{\delta}{2}}^\theta\frac{d}{dk}\right)\bigg]~.
$$
%The complete study will be given elsewhere.

\section{Conclusion}
We have used the quantum Riesz-Feller derivative in the factorization of the fractional quantum harmonic oscillator
as proposed by Olivar-Romero and Rosas-Ortiz in \cite{fff1}. We have obtained more results in analytic form as counterparts of the standard factorization of the quantum harmonic oscillator. We confirm the expressions for the fractional wavefunctions in \cite{fff1} that we obtain when the value of the asymmetry parameter is taken $\theta=1$. On the other hand, we have found that the eigenvalues have a supplementary complex term with respect to the formulas for $\theta=0$. Therefore all the `eigenstates' are metastable despite the impenetrability of the parabolic well.
A factorization with different L\'evy parameters has been also sketched up.

%However, we do not fully discard the existence of exceptional points for the time being and we hope to study this issue in more detail in the future.

\bigskip

{\bf Acknowledgement}

\noindent The organizers of the QuantFest-2019 workshop are acknowledged for the excellent conditions they offered during the event. Both authors wish to thank Dr. Oscar Rosas-Ortiz for invitation and the occasion to share memories about Bogdan Mielnik. Thanks are also due to the referee for interesting comments.

%\newpage

\bigskip
\bigskip

\noindent {\bf Appendix A: Effective calculation of $\phi_0$ for any $\alpha$}

\medskip

The calculation of $\phi_0$ proceeds as follows. The kernel equation
$$
A_{\alpha}\psi_0(x)\equiv\left(D^{\alpha/2}+x\right)\psi_0(x)=0 \qquad
%\begin{equation}\label{ap}
%\left(D^{\alpha/2}+x\right)\psi_0(x)=0
$$
%\end{equation}
%\begin{equation}\label{ap1}
is Fourier transformed by taking into account that the fractional derivative is a (quantum) Riesz-Feller derivative
$$
{\cal F}\{D^{\alpha/2}\psi_0\}+{\cal F}\{x\psi_0\}=0 \quad \longrightarrow \quad  -\Psi_{\alpha/2}^{\theta}\phi_0-i\frac{d\phi_0(k)}{dk}=0~.
$$
%\end{equation}
%Thus
%%\begin{equation}\label{ap2}
%$$
%A_{k,\alpha}=\Psi_{\alpha/2}^{\theta}+i\frac{d}{dk} \quad \longrightarrow \quad B_{k,\alpha}=\Psi_{\alpha/2}^{\theta}-i\frac{d}{dk}
%$$
%\end{equation}
Separating variables and formally integrating, we obtain
$$
\ln \phi_0=i\int \Psi_{\alpha/2}^{\theta}dk=i\int |k|^{\frac{\alpha}{2}} e^{i\,{\rm sgn}(k)\, \theta \frac{\pi}{2}}dk~.
$$
The integral in the right hand side is evaluated separately for the two possible cases:

\medskip

(i) $k>0$, then $|k|=k$, ${\rm sgn}(k)=+1$. %Then
$$
\ln\phi_0=i\int k^{\alpha/2}e^{i\theta \frac{\pi}{2}}dk=ie^{i\theta \frac{\pi}{2}}\int k^{\alpha/2}dk=i
e^{i\theta \frac{\pi}{2}}\frac{k^{\frac{\alpha}{2}+1}}{\frac{\alpha}{2}+1}~.
$$
Let $\theta=1$, then $e^{i\theta \frac{\pi}{2}}=i$, so
$$
\ln\phi_0=- \frac{k^{\frac{\alpha}{2}+1}}{\frac{\alpha}{2}+1} \quad \longrightarrow \quad \phi_0=Ce^{-\frac{k^{\frac{\alpha}{2}+1}}{\frac{\alpha}{2}+1}}~.
$$

\medskip

(ii)  $k<0$, let $k=-p$, then $p>0$ and $|p|=p$. Then
$$
\ln\phi_0=-i\int |p|^{\alpha/2}e^{i{\rm sgn}(-p)\theta \frac{\pi}{2}}dp=-ie^{-i\theta \frac{\pi}{2}}\int p^{\alpha/2}dp~.
%-ie^{-i\theta \frac{\pi}{2}}\frac{k^{\frac{\alpha}{2}+1}}{\frac{\alpha}{2}+1}
$$
For $\theta=1$:
$$
\ln\phi_0=-\int  p^{\alpha/2}dp \quad \longrightarrow \quad \phi_0=e^{-\frac{p^{\frac{\alpha}{2}+1}}{\frac{\alpha}{2}+1}}
\quad \longrightarrow \quad \phi_0=Ce^{-\frac{(-k)^{\frac{\alpha}{2}+1}}{\frac{\alpha}{2}+1}}~.
$$
From (i) and (ii), we conclude that for $\theta=1$:
%\begin{equation}\label{phi0}
$$
\phi_0=Ce^{-\frac{|k|^{\frac{\alpha}{2}+1}}{\frac{\alpha}{2}+1}}~.
$$
%\end{equation}

\bigskip
\bigskip

\noindent {\bf Appendix B: $\psi_0$ for  $\alpha=1$ and $\alpha=3/2$}\\
%%%%%%%%%%%%%%%%%%%%%%%%%%%%%%%%%%%%%%
%%%%%%%%%%%%%%%%%%%%%%%%%%%%%%%%%%%%%%

For $\alpha=1$, the $k$-space ground state wavefunction is
$$
\phi_{0}=e^{-\frac{2}{3}|k|^{3/2}}~.
$$
This is a fractional sub-Gaussian function, whose inverse Fourier transform can be written as the summation
$$
\psi_{0}=\sum_{m=0}^{2}a_{2m}x^{2m}f_{2m}(-x^{6}/6^2)~,
$$
where $f_{2m}$ are generalized hypergeometric functions which together with the coefficients $a_{2m}$ are given by
$$
f_0=\,{}_2F_{3}\left(
\frac{5}{12},\frac{11}{12}\,; \, \frac{2}{6},
\frac{3}{6},\frac{5}{6};-\frac{x^{6}}{6^2}\right)
$$
$$
a_0=2^{1/3}\cdot 3^{2/3}\cdot \Gamma\left(\frac{5}{3}\right)
$$

$$
f_2=\,{}_3F_{4}\left(
\frac{3}{4},\frac{4}{4},\frac{5}{4}\,; \, \frac{4}{6},
\frac{5}{6},\frac{7}{6},\frac{8}{6};-\frac{x^{6}}{6^2}\right)
$$
$$
a_2=-\frac 3 2
$$

$$
f_4=\,{}_2F_{3}\left(
\frac{13}{12},\frac{19}{12}\,; \, \frac{7}{6},
\frac{9}{6},\frac{10}{6};-\frac{x^{6}}{6^2}\right)
$$
$$
a_4=\frac{7}{16}\cdot \left(\frac 32 \right)^{1/3} \cdot \Gamma\left(\frac{7}{3}\right).
$$

%%%%%%%%%%%%%%%%%%%%%%%%%%%%%%%%%%%%%%
For $\alpha=3/2$, the `ground state' wavefunction in the $k$-space is the fractional sub-Gaussian function
$$
\phi_{0}=e^{-\frac{4}{7}|k|^{7/4}}~,
$$
whose inverse Fourier transform can be written as the summation
$$
\psi_{0}=\sum_{m=0}^{6}a_{2m}x^{2m}f_{2m}(-x^{14}/14^6)~,
$$
where
%$f_{2m}$ are generalized hypergeometric functions which together with the coefficients $a_{2m}$ are given next:
$$
f_0=\,{}_6F_{11}\left(
\frac{11}{56},\frac{18}{56},\frac{25}{56}, \frac{39}{56}, \frac{46}{56}, \frac{53}{56}\,; \, \frac{2}{14},
\frac{3}{14},\frac{4}{14},\frac{5}{14},\frac{6}{14},\frac{7}{14},\frac{9}{14},\frac{10}{14},\frac{11}{14},\frac{12}{14},\frac{13}{14};
-\frac{x^{14}}{14^{6}}\right)
$$
$$
a_0=\frac{7^3\cdot 7^{4/7}}{11\cdot 2\cdot 2^{1/7}\cdot 3^2\cdot 5^2}\Gamma\left(\frac{32}{7}\right)
%\cos\left(\frac{2\pi}{14}\right)\sin\left(\frac{\pi}{14}\right)\sin\left(\frac{3\pi}{14}\right)
$$
$$
f_2=\,{}_6F_{11}\left(
\frac{19}{56},\frac{26}{56},\frac{33}{56}, \frac{47}{56}, \frac{54}{56}, \frac{61}{56}\,; \, \frac{4}{14},
\frac{5}{14},\frac{6}{14},\frac{7}{14},\frac{8}{14},\frac{9}{14},\frac{11}{14},\frac{12}{14},\frac{13}{14},\frac{15}{14},\frac{16}{14};
-\frac{x^{14}}{14^{6}}\right)
$$
$$
a_2=\frac{5\cdot 2^{4/7}}{2^3\cdot 7^{11/14}}\frac{\Gamma\left(-\frac{2}{7}\right)}{\sin^2\left(\frac{\pi}{7}\right)}\frac{\sin \left(\frac{\pi}{14}\right)}{\cos\left(\frac{3\pi}{14}\right)}
$$
$$
f_4=\,{}_6F_{11}\left(
\frac{27}{56},\frac{34}{56},\frac{41}{56}, \frac{55}{56}, \frac{62}{56}, \frac{69}{56}\,; \, \frac{6}{14},
\frac{7}{14},\frac{8}{14},\frac{9}{14},\frac{10}{14},\frac{11}{14},\frac{13}{14},\frac{15}{14},\frac{16}{14},\frac{17}{14},\frac{18}{14};
-\frac{x^{14}}{14^{6}}\right)
$$
$$
a_4=\frac{13\cdot 2^{2/7}\cdot 7^{5/14}}{2^7}\frac{\Gamma\left(\frac{6}{7}\right)}{\sin^2\left(\frac{\pi}{7}\right)}\frac{\sin \left(\frac{\pi}{14}\right)}{\cos\left(\frac{3\pi}{14}\right)}
$$
$$
f_6=\,{}_7F_{12}\left(
\frac{35}{56},\frac{42}{56},\frac{49}{56}, \frac{56}{56}, \frac{63}{56}, \frac{70}{56}, \frac{77}{56}\,; \, \frac{8}{14},
\frac{9}{14},\frac{10}{14},\frac{11}{14},\frac{12}{14},\frac{13}{14},\frac{15}{14},\frac{16}{14},\frac{17}{14},\frac{18}{14},\frac{19}{14},\frac{20}{14};
-\frac{x^{14}}{14^{6}}\right)
$$
$$
a_6=-\frac{7^3\cdot 7^{1/7}}{2^8\cdot 3\cdot 5}\cot\left(\frac{\pi}{7}\right)\tan\left(\frac{\pi}{14}\right)\tan\left(\frac{3\pi}{14}\right)
$$
$$
f_8=\,{}_6F_{11}\left(
\frac{43}{56},\frac{50}{56},\frac{57}{56}, \frac{71}{56}, \frac{78}{56}, \frac{85}{56}\,; \, \frac{10}{14},
\frac{11}{14},\frac{12}{14},\frac{13}{14},\frac{15}{14},\frac{17}{14},\frac{18}{14},\frac{19}{14},\frac{20}{14},\frac{21}{14},\frac{22}{14};
-\frac{x^{14}}{14^{6}}\right)
$$
$$
a_8=\frac{7^7\cdot 7^{1/7}}{19\cdot 43\cdot 3^5\cdot 5^3\cdot 2^{14}\cdot 2^{2/7}}\Gamma\left(\frac{64}{7}\right)%\cos\left(\frac{\pi}{7}\right)
$$
$$
f_{10}=\,{}_6F_{11}\left(
\frac{51}{56},\frac{58}{56},\frac{65}{56}, \frac{79}{56}, \frac{86}{56}, \frac{93}{56}\,; \, \frac{12}{14},
\frac{13}{14},\frac{15}{14},\frac{16}{14},\frac{17}{14},\frac{19}{14},\frac{20}{14},\frac{21}{14},\frac{22}{14},\frac{23}{14},\frac{24}{14};
-\frac{x^{14}}{14^{6}}\right)
$$
$$
a_{10}=-\frac{7^8\cdot 7^{2/7}}{11\cdot 13\cdot 17\cdot 29\cdot 3^5\cdot 5^3\cdot 2^{17}\cdot 2^{4/7}}\Gamma\left(\frac{72}{7}\right)
%\cos\left(\frac{\pi}{7}\right)\sin\left(\frac{\pi}{14}\right)\sin\left(\frac{3\pi}{14}\right)
$$
$$
f_{12}=\,{}_6F_{11}\left(
\frac{59}{56},\frac{66}{56},\frac{73}{56}, \frac{87}{56}, \frac{94}{56}, \frac{101}{56}\,; \, \frac{15}{14},
\frac{16}{14},\frac{17}{14},\frac{18}{14},\frac{19}{14},\frac{21}{14},\frac{22}{14},\frac{23}{14},\frac{24}{14},\frac{25}{14},\frac{26}{14};
-\frac{x^{14}}{14^{6}}\right)
$$
$$
a_{12}=\frac{7^9\cdot 7^{3/7}}{11^2\cdot 13\cdot 59\cdot 73\cdot 3^5\cdot 5^2\cdot 2^{24}\cdot 2^{6/7}}\Gamma\left(\frac{80}{7}\right)~.
%\cos\left(\frac{\pi}{14}\right)\sin\left(\frac{\pi}{14}\right)\sin\left(\frac{3\pi}{14}\right)
$$
Counter to these expressions, the hypergeometric formula for the Gaussian function is
\begin{equation}\label{confl-g}
e^{-x^2/2}=\ _{1}F_1\left(\frac{1}{2},\frac{3}{2};-x^2/2\right)-\frac{x^2}{3}\ _{1}F_1\left(\frac{3}{2},\frac{5}{2};-x^2/2\right)~,
\end{equation}
which can be obtained from the confluent hypergeometric form of the erf function
$$
\frac{\sqrt{\pi}}{2}{\rm erf}(x)=\int^x e^{-t^2}dt=x\ _{1}F_1\left(\frac{1}{2},\frac{3}{2};-x^2\right)~,
$$
the chain rule, and the formula (see, e.g., NIST Handbook of Mathematical Functions)
$$
\frac{d}{dz}\ _{1}F_1\left(a,b;z\right)=\frac{a}{b}\ _{1}F_1\left(a+1,b+1;z\right)
$$
for $z=-x^2$.

\end{document}